\def\pmb#1{\setbox0=\hbox{#1}%
   \kern-.025em\copy0\kern-\wd0
   \kern.05em\copy0\kern-\wd0
   \kern-0.025em\raise.0433em\box0}
\def\gta{\mathrel{{\lower 3pt\hbox{$\mathchar"218$}}\hskip-8pt
   \raise 2pt\hbox{$\mathchar"13E$}}}
\def\lta{\mathrel{{\lower 3pt\hbox{$\mathchar"218$}}\hskip-8pt
   \raise 2pt\hbox{$\mathchar"13C$}}}

\def\today{\number\day\space\ifcase\month\or
  January\or February\or March\or April\or May\or June\or
  July\or August\or September\or October\or November\or December\fi
 \space\number\year}
\documentstyle[epsfig,twocolumn,aps]{revtex}

\begin{document}
\newcommand{\Vv }{{\raisebox{-1.2pt}{\makebox(0,0){$o$}}}}
\newcommand{\Zz }{{\raisebox{-1.2pt}{\makebox(0,0){$\mbox{\tiny o}$}}}}
\newcommand{\Xx }{{\special{em:moveto}}}
\newcommand{\Yy }{{\special{em:lineto}}}
\newcommand{\Ww }{{\usebox{\plotpoint}}}

\title{
\centerline{SUPERCONDUCTIVITY IN THE ``HOT SPOTS'' MODEL} 
\centerline{OF THE PSEUDOGAP STATE.}
\normalsize  \ \\ 
\centerline{\bf E.Z.Kuchinskii,\ N.A.Kuleeva,\ M.V.Sadovskii}
\vskip 0.25cm
\centerline{\it Institute for Electrophysics,\ Russian Academy of Sciences,\
Ural Branch,\ 
Ekaterinburg 620016,\ Russia.} 
\vskip 0.5cm
\begin{minipage}{6.5in}
\marginparwidth 0.625in 
\small
\baselineskip 9pt
{\bf Abstract}---We analyze the anomalies of superconducting state 
in the model of pseudogap state 
induced by fluctuations of short -- range order of ``dielectric'' (AFM (SDW) 
or CDW) type, and based on the scenario of ``hot spots'' formation on the 
Fermi surface, with the account of {\em all} Feynman
graphs for electron interaction with pseudogap fluctuations,
leading to strong scattering around the ``hot spots''. 
We determine the dependence of superconducting critical temperature $T_c$
on the effective width of the pseudogap, correlation length of short -- range
order and concentration of nonmagnetic impurities. 
We also discuss possible connection of these results with the 
general form of the phase diagram of superconducting cuprates.
\end{minipage}}
\maketitle

\setlength{\unitlength}{1in}
\makeatletter
\global\@specialpagefalse
\def\@oddhead{\footnotesize \it \hfill \ Journal of Physics and 
Chemistry of Solids}
\makeatother
\baselineskip 12pt
\normalsize \rm

One of the main problems in the physics of high -- temperature superconducting
cuprates remains the theoretical explanation of the typical phase diagram
of these compounds \cite{Lor}. Especially important is the clarification of the
nature of the pseudogap state, observed in a wide region of temperatures and
carrier concentrations \cite{MS}, which is obviously crucial for the 
understanding of electronic properties both in normal and superconducting
states. Despite the continuing discussions, the preferable ``scenario'' of
the pseudogap formation seems to be based on the model of the strong
scattering of electrons by antiferromagnetic (AFM, SDW) short -- range order
spin fluctuations \cite{MS,Pines}. 

In a recent paper \cite{KSS} we have presented microscopic derivation of
Ginzburg -- Landau expansion and studied the influence of these pseudogap
fluctuation on basic superconducting properties (for both $s-$ and $d$-wave
pairing) in the model of ``hot spots'' the Fermi surface.
Similar analysis using Gorkov's equations was given in Ref. \cite{KK}.

In the model of nearly antiferromagnetic Fermi -- liquid \cite{Sch} 
electron interaction with spin fluctuations is usually described by dynamic
susceptibility, characterized by correlation length $\xi$ and frequency
$\omega_{sf}$ of spin fluctuations, which are to be determined from
experiment and can depend both on carrier concentration and temperature. 
Both dynamic susceptibility and effective interaction are maximal (in
momentum space representation) in the vicinity of vector
${\bf Q}=(\pi/a,\pi/a)$ ($a$ -- lattice constant), which leads to the
appearance of ``two types'' of quasiparticles --- ``hot'' one with the
momenta close to the points on the Fermi surface, connected by scattering
vector $\sim{\bf Q}$, and ``cold'' one with the momenta close to the parts
of this surface surrounding diagonals of the Brillouin zone \cite{MS,Sch,KS}.

For high enough temperatures, when $2\pi T\gg \omega_{sf}$, spin dynamics
can be neglected \cite{Sch} and electron interaction with spin (pseudogap)
fluctuations reduces to scattering by appropriately defined static Gaussian
random field. For such model we can further simplify the form of effective
interaction (correlator of the random field) \cite{Sch,KS}, allowing the
complete summation of Feynman perturbation series, leading to the following
recurrence procedure determining single -- electron Green's function 
as $G(\varepsilon_n{\bf p})=G_{k=0}(\varepsilon_n{\bf p})$ from: 
\begin{equation}
G_k(\varepsilon_n{\bf p})=\frac{1}{i\varepsilon_n-\xi_{k}({\bf p})+
ikv_k\kappa-\Sigma_{k}(\varepsilon_n{\bf p})}
\label{Gk}
\end{equation}
\begin{equation}
\Sigma_k(\varepsilon_n{\bf p})=W^2s(k+1)G_{k+1}(\varepsilon_n{\bf p})
\label{Sig_k}
\end{equation}
where $\kappa=\xi^{-1}$ --- is the inverse correlation length of the pseudogap 
fluctuations, $\varepsilon_n=2\pi T(n+1/2)$, 
\begin{equation} 
\xi_k({\bf p}) 
=\left\{\begin{array}{ll} \xi_{{\bf p}+{\bf Q}} & \mbox{for odd $k$} \\ 
\xi_{\bf p} & \mbox{for even $k$}
\end{array} \right.
\label{xik}
\end{equation}
\begin{equation}
v_k=\left\{\begin{array}{ll}
|v_x({\bf p}+{\bf Q})|+|v_y({\bf p}+{\bf Q})| & \mbox{for odd $k$} \\
|v_x({\bf p})|+|v_y({\bf p})| & \mbox{for even $k$}
\end{array} \right.
\label{Vk}
\end{equation}
where ${\bf v}({\bf p})=\frac{\partial\xi_{\bf p}}{\partial {\bf p}}$ --
is the velocity of quasiparticle with the spectrum $\xi_{\bf p}$, which 
is taken in the standard form \cite{Sch}:
\begin{equation}
\xi_{\bf p}=-2t(\cos p_xa+\cos p_ya)-4t^{'}\cos p_xa\cos p_ya - \mu
\label{spectr}
\end{equation}
where $t$ -- is transfer integral between nearest, while
$t'$ -- between second nearest neighbors on the square lattice, 
$\mu$ -- is the chemical potential. Parameter $W$ with dimension of energy
determines the effective width of the pseudogap. It is clear that both $W$
and correlation length $\xi$ are (within our semiphenomenological approach) 
some functions of carrier concentration (and probably temperature) to be 
determined from the experiment \cite{Sch,KS}. 

The value of $s(k)$ is determined by combinatorics of Feynman diagrams
and for the case of Heisenberg spin (SDW) fluctuations \cite{Sch} is equal to:
\begin{equation}
s(k)=\left\{\begin{array}{cc}
\frac{k+2}{3} & \mbox{for odd $k$} \\
\frac{k}{3} & \mbox{for even $k$}
\end{array} \right.
\label{sHeis}
\end{equation}
The limits of applicability of our approximations were discussed in detail in
Refs. \cite{Sch,KS}. 

Remarkable advantage of our model is the possibility of complete summation
of {\em all} Feynman diagrams (including those with intersecting interaction
lines) also for the vertex parts, determining the response functions to an
arbitrary external perturbation. Detailed enough discussion of the vertex 
parts was presented in Ref. \cite{SS}. Here we just write down the appropriate
recurrence relations appearing after the appropriate analysis for ``triangular''
vertices in Cooper channel, analogous to those derived in Ref. \cite{KSS}
and describing the response to an arbitrary fluctuation of superconducting
order parameter (energy gap) $\Delta_{\bf q}e({\bf p})$, where the symmetry
factor determining the type (symmetry) of pairing is taken here for the
case of $d$-wave pairing as  $e({\bf p})=\cos p_xa-\cos p_ya$, and we assume
the usual case of singlet pairing. The vertex of interest to us can be written
as:
\begin{equation}
\Gamma(\varepsilon_n,-\varepsilon_n,{\bf p},{\bf -p+q})\equiv
\Gamma_{\bf p}(\varepsilon_n,-\varepsilon_n,{\bf q})e({\bf p})
\label{Gamephi}
\end{equation}
Then $\Gamma_{\bf p}(\varepsilon_n,-\varepsilon_n,{\bf q})$ is determined
by the recurrence procedure of the following form:
\begin{eqnarray}
&&\Gamma_{{\bf p}k-1}(\varepsilon_n,-\varepsilon_n,{\bf q})=1 - 
W^2r(k)G_k(\varepsilon_n,{\bf p+q})\times\nonumber\\
&&\times G_k(-\varepsilon_n,{\bf p})
\Xi_k(\varepsilon_n,{\bf p,q})
\Gamma_{{\bf p}k}(\varepsilon_n,-\varepsilon_n,{\bf q}) 
\label{Gamma} 
\end{eqnarray}
where
\begin{eqnarray}
&&\Xi_k(\varepsilon_n,{\bf p,q})=
\Biggl\{1+\nonumber\\
&&+\frac{2ik\kappa v_k}
{G^{-1}_{k}(\varepsilon_n,{\bf p+q})-G^{-1}_{k}(-\varepsilon_n,{\bf p})
-2ik\kappa v_k}\Biggr\} 
\end{eqnarray}
``Physical'' vertex is given by 
$\Gamma_{{\bf p}k=0}(\varepsilon_n,-\varepsilon_n,{\bf q})$. 
Additional combinatorial factor $r(k)$ for the case of Heisenberg spin
(SDW) fluctuations \cite{Sch} is given by:
\begin{equation}
r(k)=\left\{\begin{array}{ll}
k & \mbox{for even $k$} \\
\frac{k+2}{9} & \mbox{for odd $k$}
\end{array} \right.
\label{rk}
\end{equation}
The choice of the sign before the factor of $W^2$ in the right hand side of
Eq. (\ref{Gamma}) depends on the symmetry of superconducting order parameter
and the nature of pseudogap fluctuations \cite{KSS}, here our choice again
corresponds to the case of Heisenberg spin (SDW) fluctuations \cite{KSS2}.

\begin{figure}
\epsfxsize=7cm
\epsfysize=10cm
\epsfbox{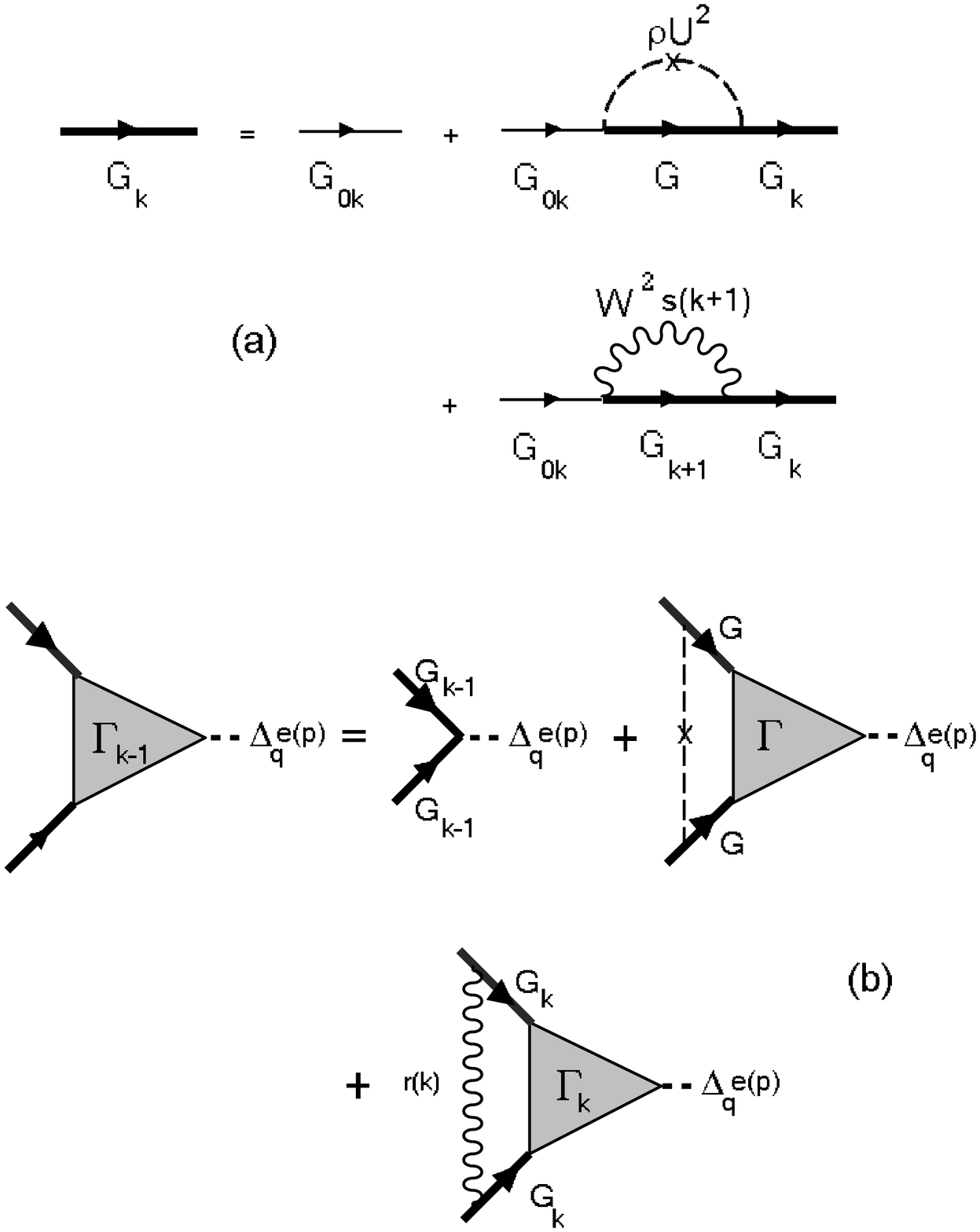}
\caption{Recurrence relations for the Green's function (a) and ``triangular''
vertex (b) with the account of scattering by random impurities.}
\label{Dysonimp}
\end{figure}

Scattering by normal (non magnetic) impurities can easily be accounted for
by self -- consistent Born approximation writing down ``Dyson equation''
for single -- electron Green's function generalizing Eq. (\ref{Gk}) and shown
diagrammatically in Fig. \ref{Dysonimp} (a), where we have introduced:
\begin{equation}
G_{0k}(\varepsilon_n{\bf p})=\frac{1}{i\varepsilon_n-\xi_k({\bf p})+
ikv_k\kappa}
\label{G0k}
\end{equation}
so that instead of (\ref{Sig_k}) we get:
\begin{equation}
\Sigma_k(\varepsilon_n{\bf p})=
\rho U^2\sum_{\bf p}G(\varepsilon_n{\bf p})+
W^2s(k+1)G_{k+1}(\varepsilon_n{\bf p})
\label{Dysrecim}
\end{equation}
where $\rho$ -- is concentration of point -- like impurities with potential
$U$, and impurity contribution to self -- energy contains the fully dressed 
Green's function $G(\varepsilon_n{\bf p})$, to be determined 
self -- consistently.
In comparison with impurity free case, we have an obvious substitution
(renormalization):
\begin{equation}
\varepsilon_n\to\varepsilon_n-\rho U^2\sum_{\bf p}ImG(\varepsilon_n{\bf p})
\label{reneps}
\end{equation}
Dropping full self -- consistency of self -- energy over impurity scattering 
we obtain just the usual substitution:
\begin{equation}
\varepsilon_n\to
\varepsilon_n+\gamma_0 sign\varepsilon_n
\label{renepsi} 
\end{equation}
where $\gamma_0=\pi\rho U^2N_0(0)$  -- is standard Born impurity scattering 
rate ($N_0(0)$ -- free electron density of states at the Fermi level).

For a ``triangular'' vertex with the account of impurity scattering we obtain
a recurrence relation of general form shown graphically in Fig.\ref{Dysonimp} 
(b). However for the case of the vertex, describing interaction with a
fluctuation of superconducting order parameter with $d$-wave symmetry this
equation is simplified considerably as the second term in the right hand
side of Fig. \ref{Dysonimp} (b) is in fact zero due to 
$\sum_{\bf p}e({\bf p})=0$ (cf. similar situation discussed in Ref. \cite{PS}). 
Then our recursion relation for the vertex part reduces to (\ref{Gamma}), 
where $G_k(\pm\varepsilon_n{\bf p})$ are determined  by Eqs. (\ref{G0k}),\ 
(\ref{Dysrecim}), i.e. are just Green's function ``dressed'' by  impurity
scattering and defined diagrammatically in Fig. \ref{Dysonimp} (a). 

Superconducting transition temperature is determined by the usual equation for
Cooper instability of the normal phase: 
\begin{equation}
1-V\chi(0;T)=0
\label{coopinst}
\end{equation}
where the generalized Cooper channel susceptibility is given by:
\begin{eqnarray}
&&\chi({\bf q};T)=-T\sum_n\sum_{\bf p}
e^2({\bf p})\times\nonumber\\
&&\times G(\varepsilon_n{\bf p+q})
G(-\varepsilon_n,-{\bf p})\Gamma_{\bf p}(\varepsilon_n,-\varepsilon_n,{\bf q})
\label{chiq}
\end{eqnarray}
Pairing interaction constant $V$ is, as usual in BCS -- like approach, 
assumed to be non zero in some layer of the width of $2\omega_c$ around the 
Fermi level and determines the ``bare'' transition temperature $T_{c0}$ in the
absence of pseudogap fluctuations by  the standard BCS equation \cite{KSS}.
Then, choosing rather arbitrarily e.g. $\omega_c=0.4t$ and $T_{c0}=0.01t$ 
we can easily find the appropriate value of $V$ corresponding to this value of
$T_{c0}$. Actually this can be easily done for any other choice of parameters
of our model.

Knowledge of Green's functions and vertices allows us to  derive 
Ginzburg -- Landau expansion in a standard way \cite{KSS}. Then, using
microscopic values of Ginzburg -- Landau coefficients we can determine all
characteristics of a superconductor close to $T_c$ (e.g. coherence length,
penetration depth, specific heat discontinuity etc.). Appropriate results will
be given in a more detailed publication \cite{KSS2}.

Numerical calculations were performed for different values of 
$t$,\ $t'$,\ and $\mu$ of the ``bare'' electronic spectrum and also for
different values of $W$ and inverse correlation length $\kappa a$.

\begin{figure}
\epsfxsize=7cm
\epsfysize=7cm
\epsfbox{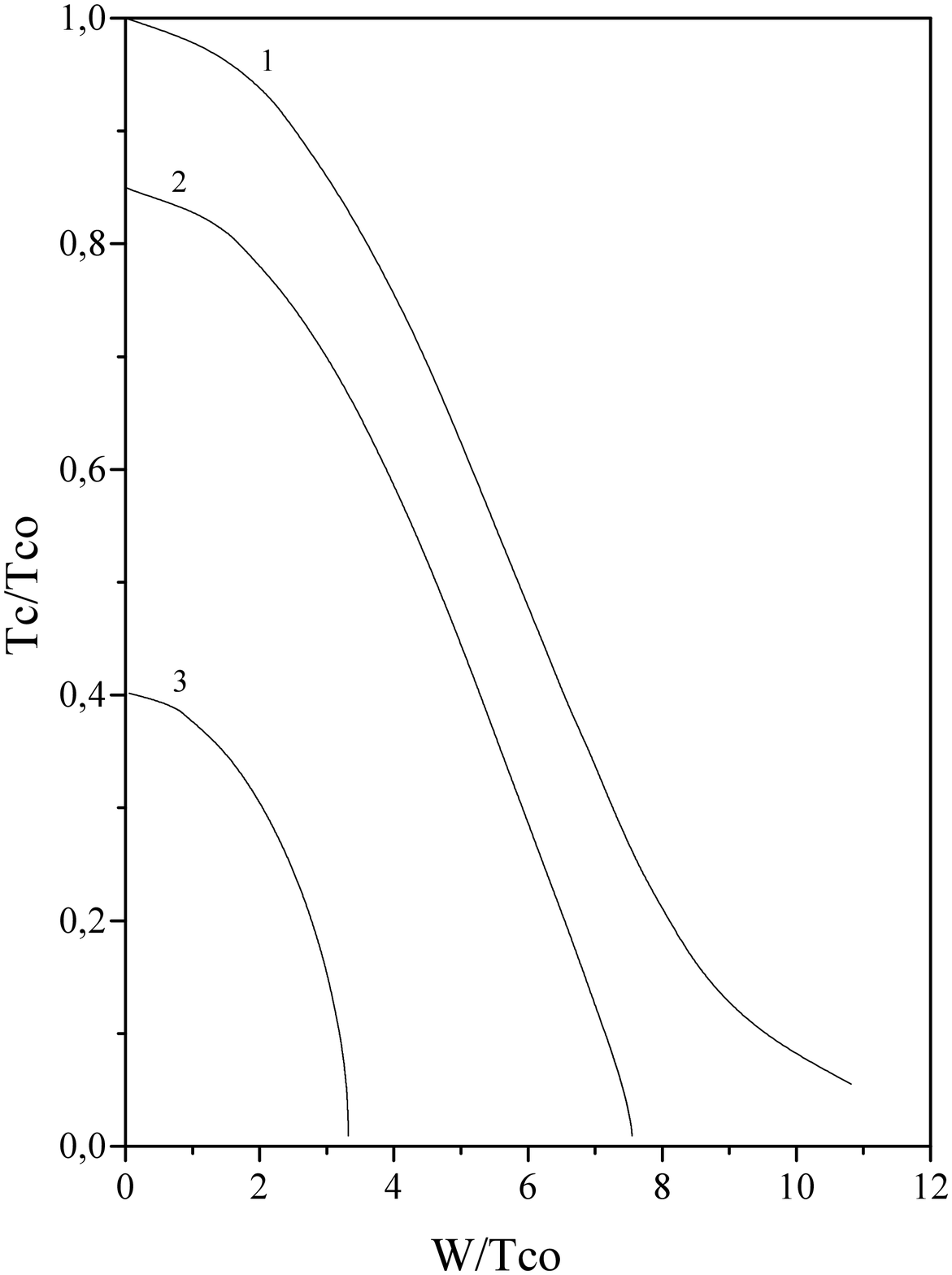}
\caption{$T_c$ dependence on the effective width of the pseudogap $W$ for the
case of $d$-wave pairing and different values of impurity scattering rate
$\gamma_0/T_{c0}$: 0 -- 1; 0.18 -- 2; 0.64 -- 3.
Inverse correlation length $\kappa a$=0.2.
} 
\label{sc1} 
\end{figure}

In Fig. \ref{sc1} we show the dependence of superconducting critical
temperature $T_c$ on the effective width of the pseudogap $W$ for different
values of impurity scattering rate. It is seen that pseudogap fluctuations
lead to significant suppression of superconductivity and that in case of
finite impurity scattering we always obtain some ``critical'' value of
$W$, corresponding to complete disappearance of $T_c$. This $T_c$ suppression
is obviously due to partial ``dielectrization'' of electronic spectrum around
the ``hot spots'' on the Fermi surface \cite{Sch,KS}. $T_c$ dependence on
the value of correlation length of short -- range order (pseudogap)
fluctuations is much slower, in all cases the growth of $\xi$ 
(diminishing $\kappa$) enhances the effect of pseudogap fluctuations
(suppression of $T_c$). We drop the appropriate results to spare space. 

\begin{figure}
\epsfxsize=7.8cm
\epsfysize=7.8cm
\epsfbox{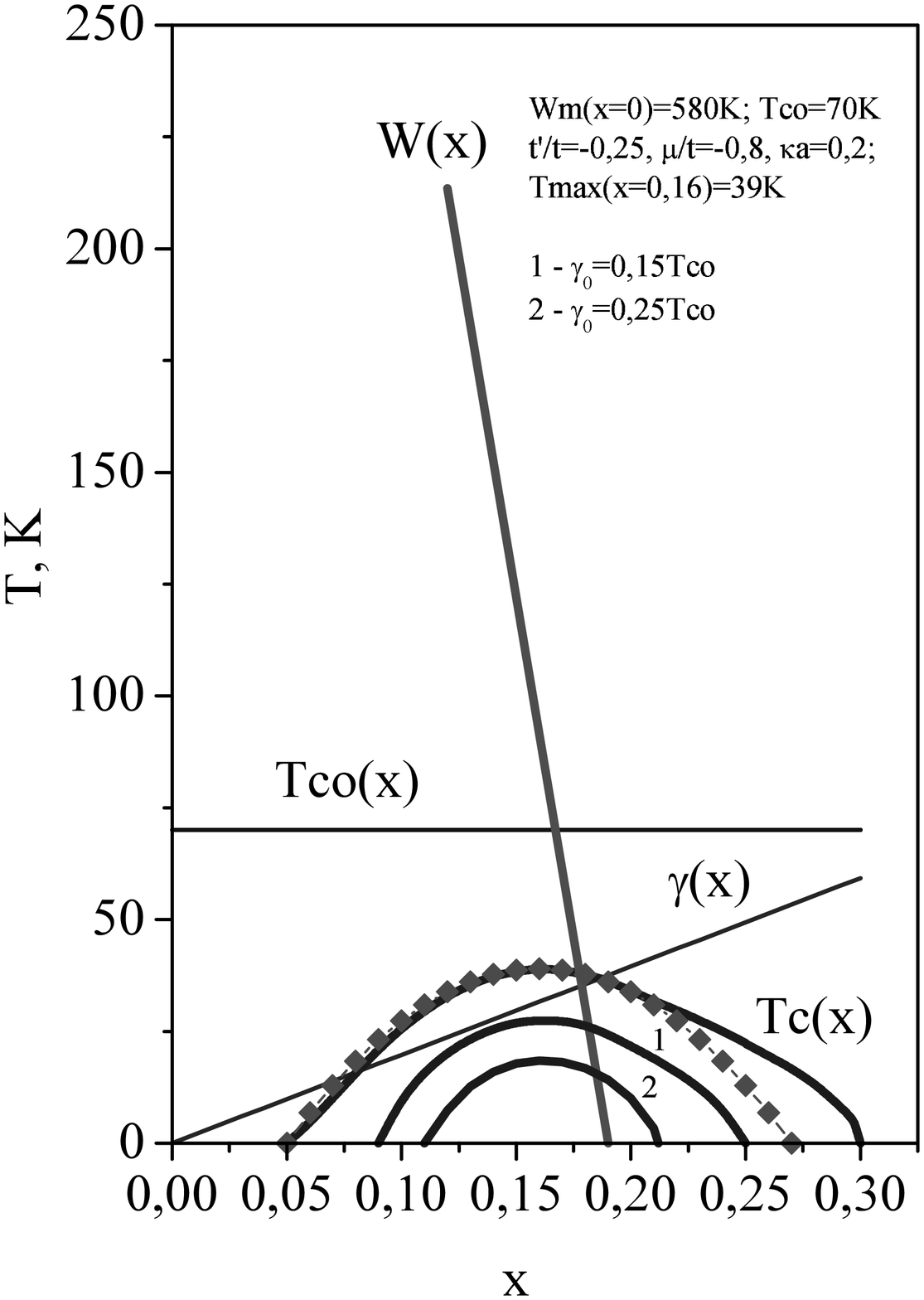}
\caption{Model phase diagram for $La_{2-x}Sr_xCuO_4$ -- type system 
in the case of
%Heisenberg (SDW) pseudogap fluctuations, 
%$d$-wave pairing and 
concentration independent ``bare'' $T_{c0}$ with ``internal'' disorder 
$\gamma(x)$ linear in $x$.}  
\label{TcSDW} 
\end{figure}

Now we can move to an attempt to model a typical phase diagram of a cuprate
superconductor within ``hot spots'' model of the pseudogap state. 
First attempt of such modelling in an oversimplified (``toy'') version of our 
model was undertaken in Ref. \cite{AC}. The main idea is to identify our
parameter $W$ with experimentally observable effective width of the pseudogap
(temperature of crossover to the pseudogap region of the phase diagram) 
$E_g\approx T^*$, which is determined from numerous experiments 
\cite{Lor,MS,Pines}. It is well known that the value of this parameter  
actually drops, more or less linearly with concentration of doping impurity
(carrier concentration), from values of the order of $10^3$K, becoming zero
at some critical concentration $x_c\approx 0.19..0.22$, slightly greater than
the ``optimal'' concentration  $x_o\approx 0.15..0.17$ \cite{Lor,NT}. 
Accordingly we just assume similar concentration dependence of our effective
pseudogap width $W(x)$. In this sense we can consider our $W(x)$ as determined
from the experiment. Then the only ``fitting'' parameter of the model is
concentration dependence of the ``bare'' superconducting transition
temperature $T_{c0}(x)$, which would have been existent in the absence of
pseudogap fluctuations. Unfortunately, as already has been noted in
Ref. \cite{KSS}, this dependence of $T_{c0}(x)$ is, in general case, unknown
and is not determined from any known experiment.
In the framework of our BCS -- like approach any significant concentration 
dependence of the ``bare'' $T_{c0}$ seems to be unrealistic\footnote{This can
be only due to some relatively smooth concentration dependence of the density
of states at the Fermi level.}. So we just assume there is no dependence of 
$T_{c0}$ on $x$ at all, but take into account the fact, that introduction of
any doping impurity obviously leads to appearance of some impurity scattering
(internal disorder), which can be described by the appropriate linear
dependence of $\gamma(x)$. Assume that it is this growth of disorder is what
leads to the complete suppression of $d$-wave pairing (according to the well
known Abrikosov -- Gorkov dependence, cf. e.g. Ref. \cite{PS}) say at 
$x=0.3$. Then our approach allows to calculate concentration dependence
$T_c(x)$ for all values of $x$. Results of such calculation with parameters
more or less appropriate for $La_{2-x}Sr_xCuO_4$, in case of Heisenberg
(SDW) pseudogap fluctuations, and with the account of this very simple model 
of impurity scattering, are shown in Fig. \ref{TcSDW}. Actual values of
parameters of the model used in these calculations are shown at the same
figure. ``Experimental'' data for $T_c(x)$ shown in this figure with 
``diamonds'' are obtained from an empirical relation \cite{NT,PT}:
\begin{equation}
\frac{T_c(x)}{T_c(x=x_o)}=1-82.6(x-x_o)^2
\label{Tcexp}
\end{equation}
which gives rather good fit to experimentally observed concentration
dependence of $T_c$ for a number of cuprates. We can see that in the whole
underdoped region our model gives practically ideal fit to ``experimental''
data, assuming quite reasonable values of $W(x)$ \footnote{Note that in 
case of Ising like pseudogap fluctuations we need rather
unrealistic values of $W(x=0)$, order a magnitude larger, than in Heisenberg
case. However, CDW like fluctuations can also give rather good fit 
\cite{KSS2}.}. 

It is interesting to analyze dependence of superconducting transition
temperature $T_c$ under additional disordering of the system at different
compositions (concentration of carriers). Such disordering was studied in a
number of experiments, e.g. introducing disorder by neutron irradiation
\cite{Gosch} or by chemical substitutions \cite{Tall}, in our model it can be
simulated by additional impurity scattering, characterized by an extra
parameter $\gamma_0$, which is just added to our parameter of an internal
disorder $\gamma(x)$. Results of this type of calculations for two values
of this additional scattering parameter are also shown in Fig. \ref{TcSDW}. 
It is seen that in complete accordance with experiments of Ref. \cite{Tall}, 
introduction of additional ``impurities'' (disorder) leads to rather fast
reduction of superconducting region on the phase diagram. Also we note that in
accordance with experiments \cite{Gosch,Tall}, suppression of superconductivity
by disorder in the underdoped region of the phase diagram (pseudogap state) is 
significantly faster, than at optimal dopings.

This work was supported in part
by the grant $N^{o}$ 02-02-16031 from the RFBR as well as by the 
RAS Programs ``Quantum macrophysics'' and ``Strongly correlated electrons
in semiconductors, metals, superconductors and magnetic materials''.

\begin{references}
\bibitem{Lor}J.L.Tallon, J.W.Loram. Physica C {\bf 349},\ 53\ (2000)
\bibitem{MS}M.V.Sadovskii. Physics Uspekhi {\bf 44},\ 515\ (2001)
\bibitem{Pines}D.Pines.\ ArXiv:\ cond-mat/0404151
\bibitem{Sch} J.Schmalian, D.Pines, B.Stojkovic.\  Phys.Rev. {\bf B60}, 667 
(1999)
\bibitem{KS}E.Z.Kuchinskii, M.V.Sadovskii. JETP {\bf 88}, 968 (1999)
\bibitem{KSS}E.Z.Kuchinskii, M.V.Sadovskii, N.A.Strigina. JETP {\bf 98}, 748 
(2004) 
\bibitem{KK}N.A.Kuleeva, E.Z.Kuchinskii. Fiz. Tverd. Tela {\bf 46}, 1557 (2004) 
\bibitem{SS}M.V.Sadovskii, N.A.Strigina. JETP {\bf 95}, 526 (2002)
\bibitem{PS}A.I.Posazhennikova, M.V.Sadovskii. JETP Letters {\bf 63}, 368
(1996);\ JETP {\bf 85}, 1162 (1997)
\bibitem{KSS2}E.Z.Kuchinskii, N.A.Kuleeva, M.V.Sadovskii -- in preparation
\bibitem{AC}A.Posazhennikova, P.Coleman. Phys.Rev. {\bf B67}, 165109 (2003)
\bibitem{NT}S.H.Naqib, J.R.Cooper, J.L.Tallon, R.S.Islam, R.A.Chakalov.
ArXiv:\ cond-mat/0312443
\bibitem{PT}M.R.Presland, J.L.Tallon, R.G.Buckley, R.S.Liu, N.E.Flower.\ 
Physica C {\bf 176}, 95 (1991)
\bibitem{Gosch}A.E.Karkin, S.A.Davydov, B.N.Goschitskii et al.\ 
%S.V.Moshkin, M.Yu.Vlasov.\ 
Fiz. Metals -- Metallogr. {\bf 76}, 103 (1993)
\bibitem{Tall}J.L.Tallon,~ C.Bernhard,~ G.V.M.Williams, J.W.Loram.
Phys.Rev.Lett. {\bf 79}, 5294 (1997)


\end{references}
\end{document}